\documentclass[prl,reprint, superscriptaddress, preprintnumbers, nofootinbib]{revtex4-2}
\usepackage[ngerman,english]{babel}
\usepackage{booktabs}
\usepackage{multirow}
\usepackage{lineno,hyperref}
\usepackage{lipsum}
\usepackage{graphicx}
\usepackage{dcolumn}
\usepackage{amsmath}
\usepackage{amsfonts}
\usepackage{natbib}
\usepackage[caption=false]{subfig}
\usepackage[range-phrase={\,\text{--}\,}, range-units={single}, separate-uncertainty=true]{siunitx}

\graphicspath{{}{}{.}}
\modulolinenumbers[5] 

\begin{document}

\title{Improved $S$-factor of the $^{13}$C(p,$\gamma$)$^{14}$N reaction at $E$\textsubscript{p}$\,=\,$330-740~keV and parameters of resonances at 448~keV and 551~keV}

\author{J. Skowronski}\affiliation{Universit\`a degli Studi di Padova, 35131 Padova, Italy} \affiliation{INFN, Sezione di Padova, 35131 Padova, Italy}

\author{E. Masha}\affiliation{Helmholtz-Zentrum Dresden-Rossendorf, 01328 Dresden, Germany}

\author{D.Piatti}\email{denise.piatti@pd.infn.it}\affiliation{Universit\`a degli Studi di Padova, 35131 Padova, Italy} \affiliation{INFN, Sezione di Padova, 35131 Padova, Italy}

\author{M. Aliotta}\affiliation{SUPA, School of Physics and Astronomy, University of Edinburgh, EH9 3FD Edinburgh, United Kingdom}

\author{D. Bemmerer}\affiliation{Helmholtz-Zentrum Dresden-Rossendorf, 01328 Dresden, Germany}

\author{A. Boeltzig}\affiliation{Helmholtz-Zentrum Dresden-Rossendorf, 01328 Dresden, Germany}

\author{A. Caciolli}\affiliation{Universit\`a degli Studi di Padova, 35131 Padova, Italy} \affiliation{INFN, Sezione di Padova, 35131 Padova, Italy}

\author{F. Cavanna}\affiliation{INFN, Sezione di Torino, 10125 Torino, Italy}

\author{L. Csedreki}\affiliation{HUN-REN, Institute for Nuclear Research (HUN-REN Atomki), PO Box 51, 4001 Debrecen, Hungary}

\author{R. Depalo}\affiliation{Universit\`a degli Studi di Milano, 20133 Milano, Italy}\affiliation{INFN, Sezione di Milano, 20133 Milano, Italy}

\author{P. Hempel}\affiliation{Helmholtz-Zentrum Dresden-Rossendorf, 01328 Dresden, Germany}

\author{M. Hilz}\affiliation{Helmholtz-Zentrum Dresden-Rossendorf, 01328 Dresden, Germany}

\author{G. Imbriani}\affiliation{Universit\`a degli Studi di Napoli “Federico II”, 80126 Napoli, Italy}\affiliation{INFN, Sezione di Napoli, 80126 Napoli, Italy}

\author{T. Lossin}\affiliation{Helmholtz-Zentrum Dresden-Rossendorf, 01328 Dresden, Germany}

\author{M. Osswald}\affiliation{Technische Universität Dresden, Zellescher Weg 19, 01069 Dresden, Germany}

\author{B. Poser}\affiliation{Helmholtz-Zentrum Dresden-Rossendorf, 01328 Dresden, Germany}

\author{D. Rapagnani}\affiliation{Universit\`a degli Studi di Napoli “Federico II”, 80126 Napoli, Italy}\affiliation{INFN, Sezione di Napoli, 80126 Napoli, Italy}

\author{S. Rümmler}\affiliation{Helmholtz-Zentrum Dresden-Rossendorf, 01328 Dresden, Germany}

\author{K. Schmidt}\affiliation{Helmholtz-Zentrum Dresden-Rossendorf, 01328 Dresden, Germany}

\author{R.~S. Sidhu}\affiliation{SUPA, School of Physics and Astronomy, University of Edinburgh, EH9 3FD Edinburgh, United Kingdom}\affiliation{School of Mathematics and Physics, University of Surrey, Guildford, GU2 7XH, United Kingdom}

\author{T. Sz\"ucs}\affiliation{HUN-REN, Institute for Nuclear Research (HUN-REN Atomki), PO Box 51, 4001 Debrecen, Hungary}

\author{A. Tóth}\affiliation{HUN-REN, Institute for Nuclear Research (HUN-REN Atomki), PO Box 51, 4001 Debrecen, Hungary}\affiliation{University of Debrecen, Doctoral School of Physics, Egyetem tér 1, H-4032 Debrecen, Hungary}

\author{S. Turkat}\affiliation{Universit\`a degli Studi di Padova, 35131 Padova, Italy}\affiliation{INFN, Sezione di Padova, 35131 Padova, Italy}\affiliation{Technische Universität Dresden, Zellescher Weg 19, 01069 Dresden, Germany}

\author{S. Vincent}\affiliation{Helmholtz-Zentrum Dresden-Rossendorf, 01328 Dresden, Germany}

\author{S. Werner}\affiliation{Helmholtz-Zentrum Dresden-Rossendorf, 01328 Dresden, Germany}

\author{A. Yadav}\affiliation{Helmholtz-Zentrum Dresden-Rossendorf, 01328 Dresden, Germany}

\begin{abstract}
The $^{13}$C(p,$\gamma$)$^{14}$N reaction is the second reaction of the CNO cycle. This cycle takes place in our Sun and fuels massive, Red, and Asymptotic Giant Branch stars. The $^{13}$C(p,$\gamma$)$^{14}$N rate affects the final abundances of $^{12,13}$C and $^{19}$F nuclides, with impact on our understanding of the i- and s-process, giant star nucleosynthesis and mixing processes, and ultimately the chemical evolution of the Galaxy. 
Here, we report on a new measurement of the $^{13}$C(p,$\gamma$)$^{14}$N cross-section, which has been performed at the Felsenkeller shallow-underground laboratory in Dresden (Germany).
The present $S$-factor results agree at low energy with LUNA data but are about 20\% lower than previous literature data over the whole energy range explored, $E\,=\,$310~-~680 keV. The narrow resonance corresponding to the 7966.9(5) keV excited state has been investigated and we report a new resonance strength, $\omega \gamma\,=\,$18(2)~meV. 
In addition a new R-matrix fit is presented, from which new parameters for the broad resonance corresponding to the 8062.0(10) keV excited state are derived and a new extrapolation for the total $S$-factor down to zero energy is obtained, $S$\textsubscript{tot}(0) = 6.4(4) keV b. Finally a new reaction rate is calculated and reported here.
\end{abstract}

\maketitle

\section{\label{sec:intro}Introduction}

The $^{13}$C(p,$\gamma$)$^{14}$N reaction ($Q$\textsubscript{value} = 7550.56356(22)~keV \cite{Wang21-CPC}) is the second radiative capture reaction in the normal or cold CNO cycle, $T\,=\,$0.02 - 0.1~GK \cite{Iliadis07-Book}:

\begin{equation}
\begin{split}
 &^{12}{\rm C}(p,\gamma)^{13}{\rm N}(\beta ^{+}\,\nu)^{13}{\rm C}(p,\gamma)^{14}{\rm N}\xrightarrow{}\\
 &\xrightarrow{}^{14}\!\!{\rm
 N}(p,\gamma)^{15}{\rm O}(\beta ^{+}\,\nu)^{15}{\rm N}(p,\alpha)^{12}{\rm C} , 
\end{split}
\end{equation}
which is the main nuclear energy source in massive stars during main sequence phase, and in stars in more advanced stages, \textit{i.e.} Red Giant Branch (RGB) and Asymptotic Giant Branch (AGB) stars \cite{Rolfs88-Book}.

RGB and AGB stars are believed to significantly contribute to Galactic chemical evolution \cite{Romano-2017}. For instance, AGB stars produce fluorine \cite{Jorissen-1992}, and may play a key role in the long standing puzzle of the Galactic fluorine abundance \cite{Renda-2004, Womack-2023}. The proposed mechanism to explain the $^{19}$F production in AGB stars, via the $^{18}$O(p, $\alpha$)$^{15}$N($\alpha$, $\gamma$)$^{19}$F reaction chain, strongly depends on the efficiency of proton ingestion in the top layer of the He-intershell and on the reaction rate of a number of reactions \cite{Lugaro-2004}. Among these reactions the $^{13}$C(p,$\gamma$)$^{14}$N affects both the formation of $^{13}$C in the partial mixing zone, source of neutrons via the $^{13}$C($\alpha$,n)$^{16}$O reaction (recently measured in the Gamow window \cite{Ciani21-PRL, Gao22-PRL}), and the production of $^{14}$N from which $^{18}$O is synthesized via either $^{14}$N(n, p)$^{14}$C($\alpha$, $\gamma$)$^{18}$O or $^{14}$N($\alpha$, $\gamma$)$^{18}$F($\beta^{+}$, $\nu$)$^{18}$O reaction chain \cite{Lugaro-2004}. In particular, a lower rate for the $^{13}$C(p,$\gamma$)$^{14}$N reaction than reported in \cite{king} would increase significantly the fluorine output \cite{Lugaro-2004}.

Proton ingestion at the top and deeper into the He inter-shell of low to intermediate mass stars is also crucial for s- and i-processes \cite{Gallino1998, Choplin-2021, Cristallo-2009}. The local p/$^{12}$C ratio as well as the $^{12,13}$C(p,$\gamma$)$^{13,14}$N reaction rates are crucial for the formation of the $^{13}$C pocket \cite{Lugaro-2003, Cristallo-2009}. At high p/$^{12}$C ratio most of the $^{13}$C might be converted into $^{14}$N, a major neutron poison, with significant impact on the s- and i- process nucleosynthesis \cite{Lugaro-2001, Lugaro-2003}.

Moreover, the $^{12,13}$C(p,$\gamma$)$^{13,14}$N reaction rates are crucial for the determination of the carbon isotopic ratio observed in the solar system \cite{Scott06-AA, Ayres-2013}, in the interstellar medium \cite{milam2005}, in presolar grains \cite{Zinner2014}, and in metal poor stars \cite{Spite-2006, Christlieb2002, Aguado-2023}.
The observed $^{12}$C/$^{13}$C ratio combined with precise knowledge of  $^{12,13}$C(p,$\gamma$)$^{13,14}$N reaction cross sections would be a robust diagnostic of deep mixing \cite{Spite-2006}, the early Galactic evolution and the Galaxy chemical evolution \cite{Christlieb2002, milam2005, Romano-2017, Molaro2023}.

Many experiments have been dedicated to measure the $^{13}$C(p,$\gamma$)$^{14}$N reaction cross-section \cite{hester, vogl, king, genard, Skowronski-2023}.
At the energy range of astrophysical interest, $E$\footnote{In the following $E$\textsubscript{p} and $E$ will denote the proton energy in the laboratory and in the center of mass frame, respectively.}$\,\le\,120\,$keV, few data are available \cite{hester, vogl, king, Skowronski-2023}, all obtained in direct kinematic measurements. Furthermore, only Ref. \cite{king} reported results for all known transitions. The cross sections data for the main transition, namely to the ground state, scatter by up to 20\% and have uncertainties by up to 10\%. Moreover, a recent measurement performed by the LUNA collaboration reported a 30\% lower cross section with respect to the literature at 70$\,<E<\,$370~keV \cite{Skowronski-2023}.

Available cross section data at higher energies are dominated by the $E$\textsubscript{p} = 551(1)~keV \cite{Ajzenberg-Selove_1991NuPhA} broad resonance. While the main datasets \cite{vogl, king, genard} qualitatively agree, the resonance parameters, crucial for extrapolation down to low energies, are, indeed, affected by high uncertainties, since conflicting results are reported \cite{vogl, king, genard}, see Tab. \ref{tab:table1}. On the other hand, the narrow resonance at $E$\textsubscript{p} = 448.5(5) keV \cite{vogl} is poorly constrained by experimental data, as reported in Tab. \ref{tab:table1}. 

The present work re-investigates the $^{13}$C(p,$\gamma$)$^{14}$N reaction cross section in the energy range $E$ = 310$\,-\,$680~keV. The experiment was performed at the shallow underground accelerator facility at Felsenkeller, Dresden (Germany) \cite{Bemmerer25-EPJA}.

\begin{table*}[htbp]
    \caption{The $E$\textsubscript{x}$\,=\,$\SI{8062}{\kilo\electronvolt} and \SI{7966.9}{\kilo\electronvolt} resonance parameters from present R-matrix fit and from literature. The resonance energy, $E_{r}$, is provided in the laboratory system. Total and partial widths, $\Gamma$\textsubscript{tot, p, $\gamma$}, are provided in the center of mass. For all the known transitions the radiative width is reported.}
    \centering
    \begin{tabular}{@{}clllllllll@{}}
        \midrule
        \midrule
        Reference & $E_{r}$ (keV)  & $\Gamma$\textsubscript{tot} (keV) & $\Gamma$\textsubscript{p} (keV) & \multicolumn{5}{c}{$\Gamma _{\gamma}$ (eV)} \\
        & & & & $\rightarrow 0~\textup{keV}$ & $\rightarrow 2312~\textup{keV}$ & $\rightarrow 3948~\textup{keV}$ & $\rightarrow 4915~\textup{keV}$ & $\rightarrow 5105~\textup{keV}$ & $\rightarrow 5691~\textup{keV}$ \\
        \midrule
         This work & $556.4(8)$ & & $34.9(4)$ & $7.2(7)$ & $0.13(1)$ & $1.3(1)$ & $0.16(2)$ & $0.040(6)$ & $0.41(4)$ \\
         \cite{chakraborty2015} & $557(3)$ & & $37.2(3)$ & $9.09(5)$ & $0.2(4)$ & $1.544(9)$ & $0.26(1)$ & $0.074(8)$ & $0.612(6)$ \\
         \cite{Li-2012} & 557.4(22) & & 37.4(31) & 9.1(6) & 0.22(5) & 1.57(7) & 0.27(3) & 0.074(19) & 0.61(8) \\
        \cite{Artemov-2008} & 555.7 & & 37.14 & 9.20 & 0.22 & 1.53 & 0.265 & 0.077 & 0.60 \\
        \cite{Mukhamedzhanov-2003} & 556.8 & & &9.10 & 0.22 & 1.53 & 0.260 & 0.085 & 0.63 \\
        \cite{king} & 558(1) & 37.1(9) & & & & & & & \\
        \cite{Ajzenberg-Selove_1991NuPhA} & 551(1) & 23(1)& & 9.9(25)&0.17(5) &1.56(40) & 0.23(6) & 0.03(2) & 0.43(12)\\
       \cite{vogl} & 555.0(15) & 38.2 & & & & & & & \\
      \cite{genard} & 550.6(5) & 37.2(12) & & & & & & & \\
      \midrule
      \cite{king} & 450.4(5) &  & & & & & & & \\
      \cite{Ajzenberg-Selove_1991NuPhA} & 448.5(5) & $<0.37$ & & & & & & & \\
      \cite{vogl} & 448.5(5) & $<\,0.4$ & & & & & & & \\
        \midrule
        \midrule
    \end{tabular}
    \label{tab:table1}
\end{table*}

\section{\label{sec:setup}Experimental setup}
\begin{figure}[ht!]
\centering
\includegraphics[width=0.6\textwidth]{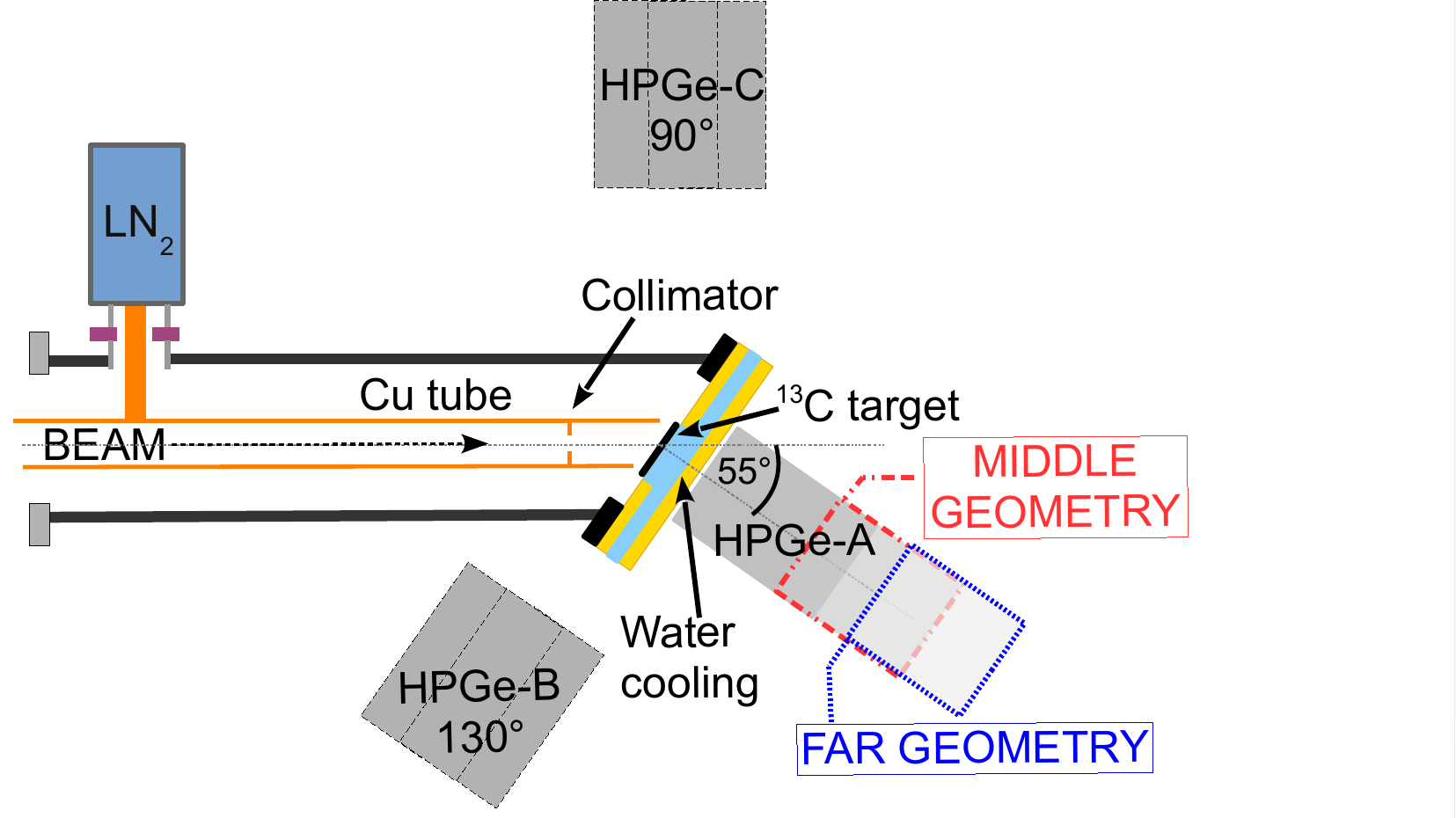}
\caption{Schematic top view of the present setup (not to scale). The detector main features are in Tab.~\ref{tab:table2}. Detector A was positioned in close, middle and far geometry for the cross-section, the broad resonance and the angular distribution investigation, respectively.}
\label{fig:figure1}
\end{figure}

A schematic view of the experimental setup is shown in Fig. \ref{fig:figure1}. The 5$\,$MV Pelletron accelerator of Felsenkeller Laboratory, in internal source mode, provided molecular H\textsubscript{2}$^{+}$ beam with typical intensity of $\sim\,10\,$p$\mu$A and energies ranging between $E$\textsubscript{p} = 330$\,-\,$750 keV. The accelerator energy calibration used during the experiment is detailed in Ref. \cite{Bemmerer25-EPJA}, which reports deviations by at most 0.15\% after an in-tank maintenance and an uncertainty of about $0.8$~keV at 500~keV.

The beam was analyzed by a magnetic analyzer and collimated by three apertures along the beamline. An additional collimator, elliptical 15$\times$7.3~mm, was installed in a copper tube, 310~mm long and with a diameter of 22~mm, positioned at 3 mm from the target. The Cu tube was in thermal contact with LN\textsubscript{2} to improve local vacuum conditions and prevent carbon build-up on the target.  The Cu tube was biased at a voltage of $-1000$~V, for secondary electron suppression. To ensure a complete suppression of secondary electrons a permanent magnet was added just below the target location, in a position that maximized secondary electron suppression, according to dedicated tests.

Two targets, namely 1 and 2, were used during the measurement, both produced at ATOMKI by evaporation of enriched carbon powder (99\% nominal $^{13}$C enrichment level and 99.8\% nominal chemical purity from ADVENT) on Ta disks (0.25 mm thick and with diameter of 27 mm from Goodfellow) previously cleaned both mechanically and chemically \cite{Ciani20-EPJA}.

The target was mounted at 55$^\circ$ with respect to the beam direction on a holder made of aluminum and stainless steel. To prevent degradation, the target backing was in direct contact with water at a set temperature of 16$^\circ$C. The target holder was electrically insulated from the beam line and served as a Faraday cup for beam current measurements.

\begin{figure}[htbp]
    \centering
        \includegraphics[width=0.49\textwidth]{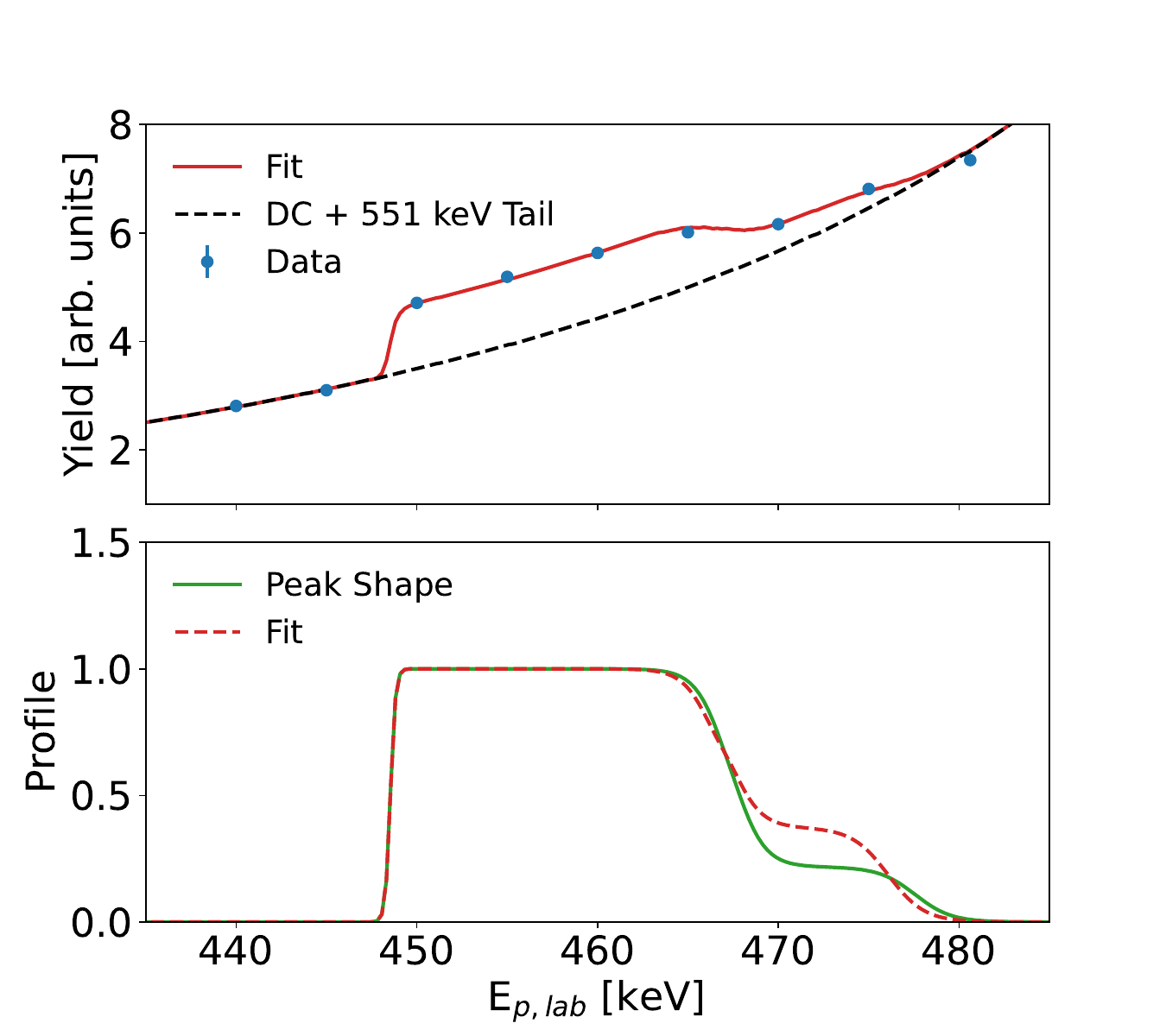}
    \caption{Top: The $E$\textsubscript{p}$\,=\,$\SI{448.5}{\kilo\electronvolt} resonance scan fit with target 1, blue dots and red solid line. The direct capture and tail of the broad resonance yield was calculated with eq. \ref{eq_1} and R-matrix curve from Ref.\cite{Skowronski-2023} is shown by dashed black line. Bottom: A comparison between the target profile obtained via peak-shape analysis at the resonance energy, green solid line and via 448.5 keV resonance scan (after subtraction of the direct capture and broad resonance contribution), red dashed line. Both profile are normalized to 1 for clarity. The target thickness is 18.4(4) keV and 18.7(4) keV from peak-shape and resonance analysis, respectively.}
    \label{fig:figure2}
\end{figure}

In-situ peak-shape analysis was performed to characterize the targets and to monitor their degradation under beam irradiation~\cite{Ciani20-EPJA}.
Periodic runs were performed at $E$\textsubscript{p} = 500 keV and the $^{13}$C(p,$\gamma$)$^{14}$N $\gamma$-ray, corresponding to the transition to the ground state, was analyzed. To check the robustness of the method, the peak-shape analysis was performed also on additional primary $\gamma$-rays, and results were in excellent agreement. In addition, for target 1 the narrow resonance at $E$\textsubscript{p} = 448.5 keV was scanned and fitted to cross check the peak shape analysis result, see Fig. \ref{fig:figure2}.  No significant degradation was observed in both targets and the obtained target thicknesses at \SI{500}{\kilo\electronvolt} were \SI{17.7 \pm 1}{\kilo\electronvolt} and \SI{14.5 \pm 0.6}{\kilo\electronvolt} for target 1 and 2, respectively.

\begin{figure*}[ht!]
    \centering
    \includegraphics[width=0.99\textwidth]{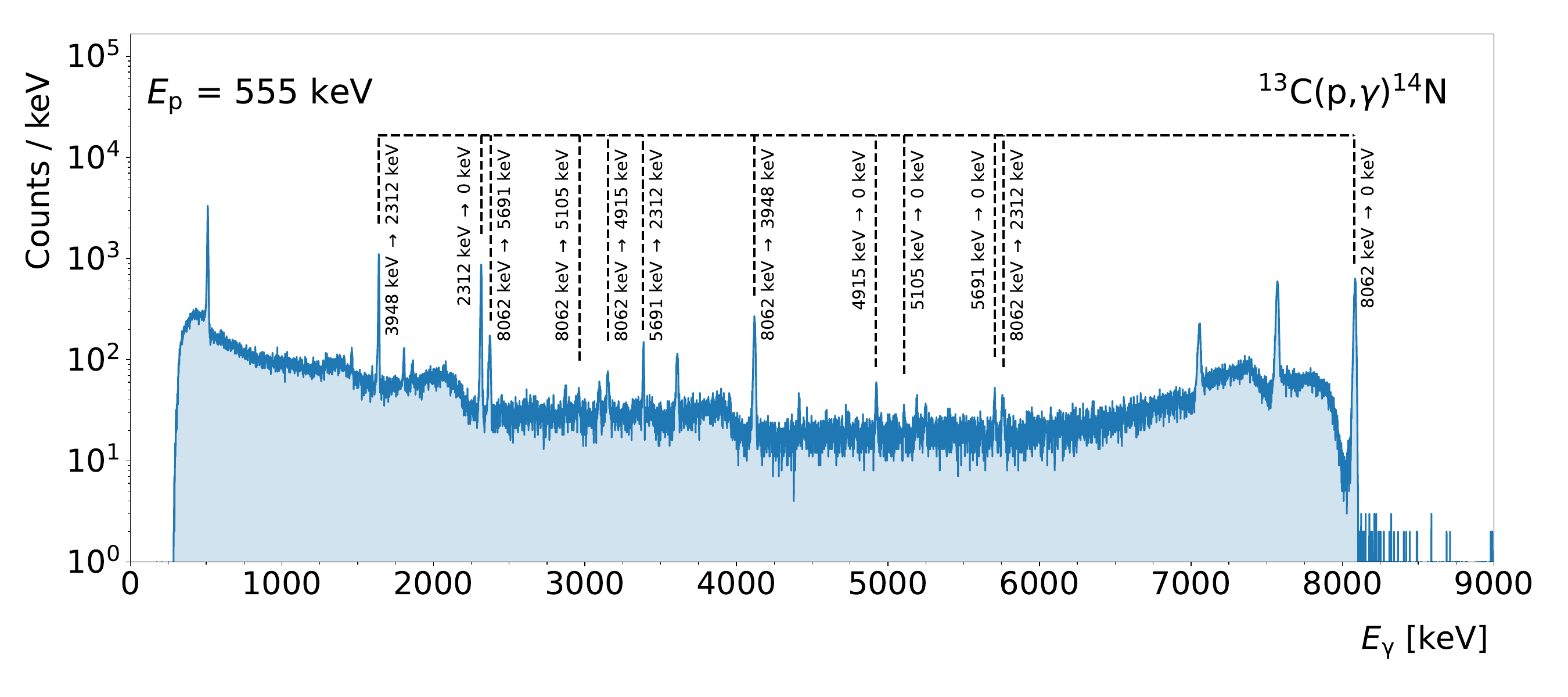}
    \caption{The $\gamma$-ray spectrum acquired at $E$\textsubscript{p} = 555~keV ($E\,=\,$515~keV) with detector A in close geometry. The main $\gamma$-rays from the $^{13}$C(p,$\gamma$)$^{14}$N reaction are shown with black dashed lines. Other prominent peaks not marked are the escape peaks and the annihilation peak.}
    \label{fig:figure3}
\end{figure*}

In the energy range investigated here the $^{13}$C(p,$\gamma$)$^{14}$N reaction emits many $\gamma$-rays, corresponding to transitions to excited states in the compound nucleus and ranging between $2312\,$keV$\,\le\,E_{\gamma}\,\le\,$($Q$\textsubscript{value} + $E$), see a typical spectrum in Fig. \ref{fig:figure3}. The $\gamma$ rays were detected with an HPGe detector, labeled A in Tab. \ref{tab:table2} and Fig. \ref{fig:figure1}, positioned at 55(2)$^\circ$ with respect to the beam direction and at 3.5(2)~cm from the center of the target. To keep the dead time below 2\% the detector A was moved to 10.0(2)~cm distance from target, middle geometry in Fig. \ref{fig:figure1}, when investigating the 551 keV resonance.
Moreover the detector A and two additional cluster detectors, B and C in Tab. \ref{tab:table2} and in Fig. \ref{fig:figure1}, were positioned in far geometry, at 20, 21.6 and 36~cm from the target, respectively, to check the angular distribution.

\begin{table}[]
    \centering
    \caption{HPGe detectors parameters: type, relative efficiency, angle to the beam direction (uncertainty 2 deg), and distance from detector end-cap to target center (uncertainty 0.2 cm). Detector A was in close geometry for the cross-section measurement, at 10 cm, middle geometry, on top of the broad resonance to keep the dead time below 2\% and at 20 cm, far geometry, for angular distributions investigation. }
    \begin{tabular}{c c c c r}
    \midrule
    \midrule
        Detector & Cluster & Relative Efficiency & Angle & Distance \\
        & Type & [\%] & [deg] & [cm] \\
        \midrule
        A & single & 100 & 55 & 3.5/10.0/20.0 \\
        B & 3 crystals & 3$\times$60 & 130 & 21.6 \\
        C & 3 crystals & 3$\times$60 & 91 & 36.5\\
        \midrule
        \midrule
    \end{tabular}
    \label{tab:table2}
\end{table}

\section{\label{sec:analysis}Analysis}
The self-triggered data acquisition was in list mode for all detectors. For each event both energy and time were recorded, and two additional channels were dedicated to the acquisition of the currents on the Cu tube and on the target.

The $^{13}$C(p,$\gamma$)$^{14}$N experimental yield, $Y$, was obtained for each run and each transition, $\gamma_i$, as follows:

\begin{equation}
Y_i = \frac{N_{\gamma_i}}{N_p \cdot \xi_{\gamma_i} \cdot W_i(\theta)}
\label{eq_1}
\end{equation}

The net counts $N_{\gamma_i}$ were obtained for detector A with typical statistical uncertainty of 1-2\%. The number of impinging protons, $N_p$, was derived from the accumulated charge multiplied by 2 due to the fact that the beam was molecular. The uncertainty of the charge collection is conservatively estimated at 3\%, to take into account imperfections in the Faraday cup \cite{Skowronski-2023b}.

The absolute full-energy peak efficiency of all detectors was measured using point-like radioactive sources ($^{137}$Cs, $^{60}$Co, and $^{88}$Y), with activities calibrated by Physikalisch-Technische Bundesanstalt (PTB) to 1\% accuracy. The efficiency data were extended up to 10 MeV using the well-known $^{14}$N(\textit{p},$\gamma$)$^{15}$O resonance at $E$\textsubscript{p} = 278~keV \cite{SFIII} and the $^{27}$Al(\textit{p},$\gamma$)$^{28}$Si resonance at $E$\textsubscript{p} = 992~keV \cite{
Harissopulos-2000}. For detector A, in all configurations in Tab.\ref{tab:table2}, the full-energy efficiency, $\xi_{\gamma_i}$, at the $E_{\gamma_i}$ of interest was obtained through an analytic fit, accounting for the True Coincidence Summing, as described in Ref. \cite{Imbriani_2005EPJA}.
The efficiency uncertainty was estimated using the Markov Chain Monte Carlo approach with the $emcee$ routine \cite{emcee}. The relative uncertainty is about \SIrange{4}{5}{\percent} through all the energy range of the $^{13}$C(p,$\gamma$)$^{14}$N reaction $\gamma$-rays.
After dismounting the target, a clear off-centered beamspot was observed. The effect of the beamspot position on the efficiency of the detector A in close geometry was estimated to be of 11\% via Geant4 \cite{Agostinelli_2003} simulation. This correction was applied to the efficiency. However, since no online monitor of the beamspot position was available, we estimated the total systematic uncertainty of the efficiency to be \SI{10}{\percent}.

Summing effects were negligible for detectors B and C because of their large distance from the target, thus their $\xi_{\gamma_i}$ were obtained via simple fit of the calibration data as reported in Ref. \cite{Gilmore08-Book}. Additionally, the effect of the beamspot position on these detectors, as well as on detector A in far geometry, was also checked through Monte Carlo simulations and it was found to be negligible. The total uncertainty on efficiency of detectors B, C and A (at 20 cm) is assumed to be \SI{5}{\percent} from the fit.
The efficiency was required for detectors B, and C to investigate the angular distribution of the different transition observed in the $^{13}$C(p,$\gamma$)$^{14}$N reaction.

\begin{figure*}[htbp]
    \centering
   \begin{minipage}[b]{.22\textwidth}
    \centering
        \includegraphics[width=\textwidth]{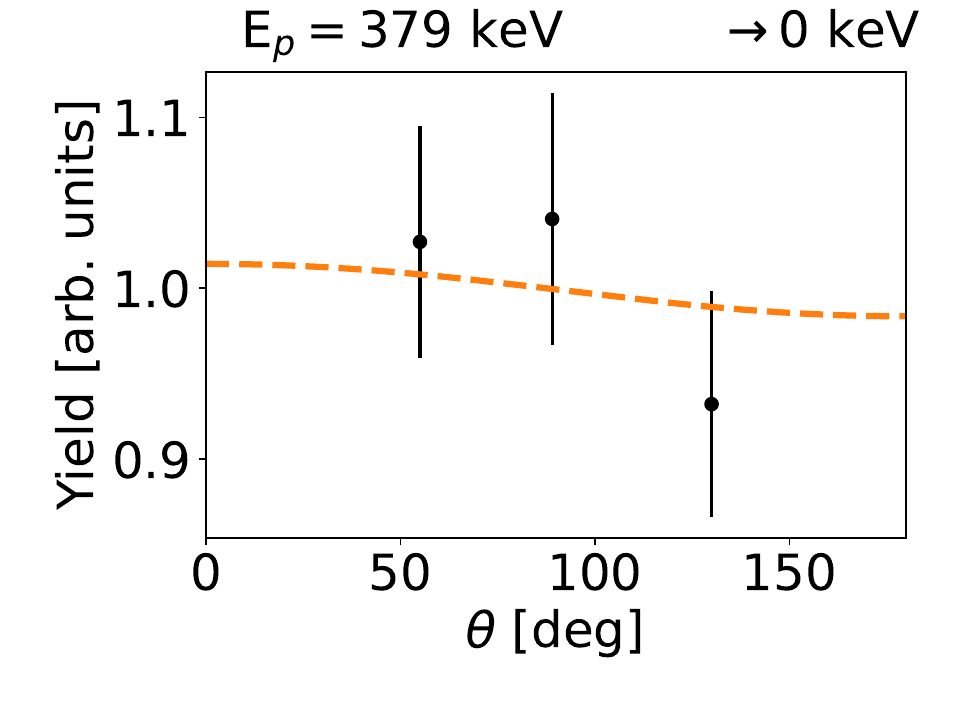}
    \end{minipage}
    \hfill
   \begin{minipage}[b]{.22\textwidth}
    \centering
        \includegraphics[width=\textwidth]{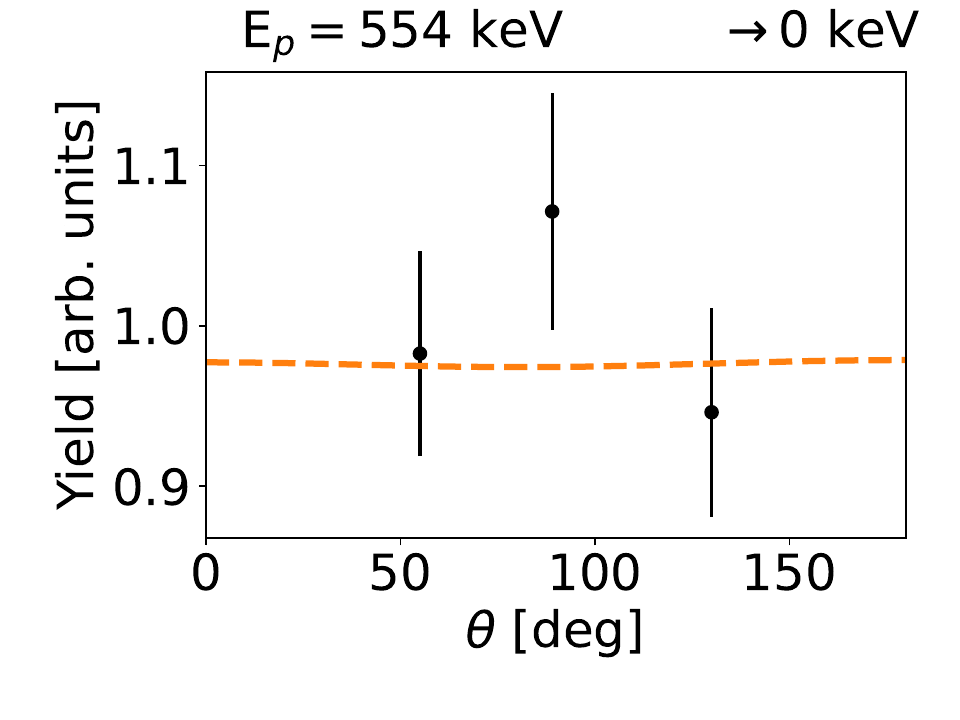}
    \end{minipage}
    \hfill
     \begin{minipage}[b]{.22\textwidth}
    \centering
        \includegraphics[width=\textwidth]{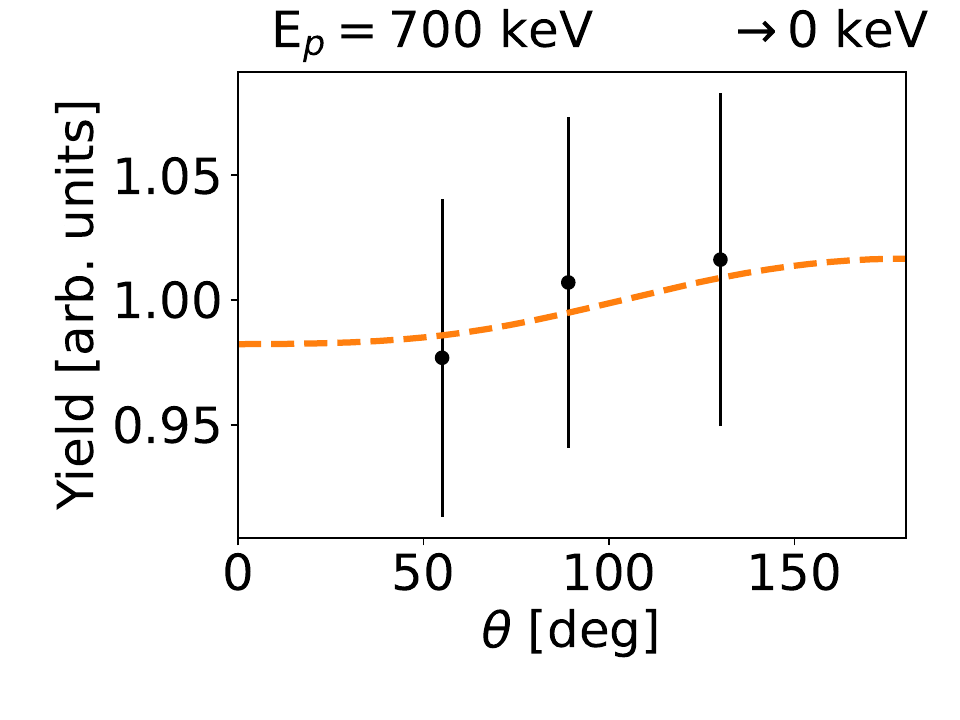}
    \end{minipage}
    \hfill
    \begin{minipage}[b]{.22\textwidth}
    \centering
        \includegraphics[width=\textwidth]{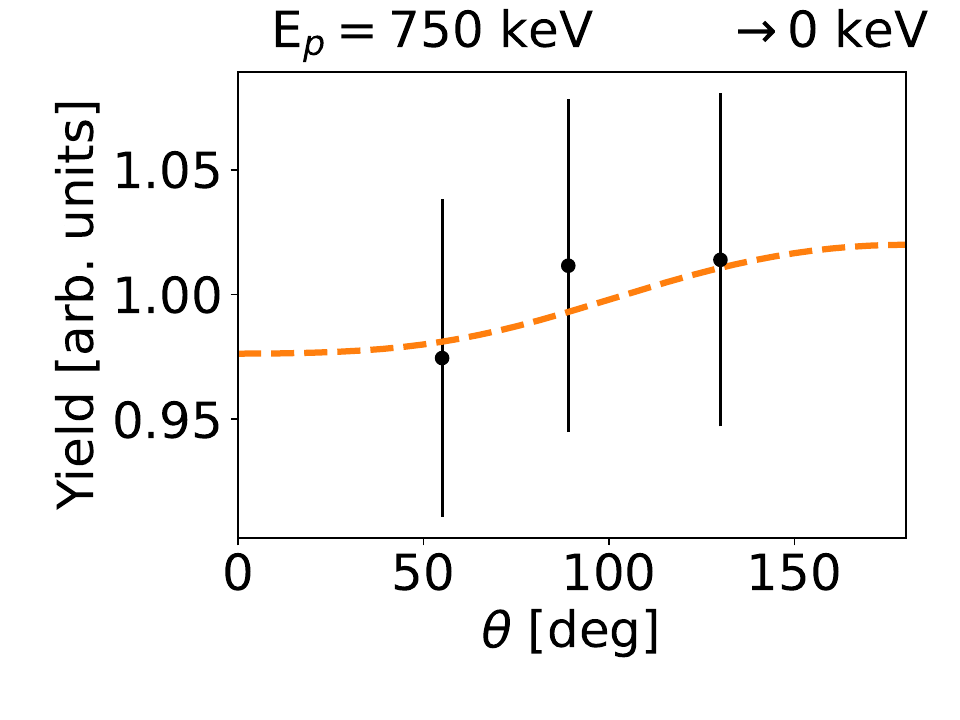}
    \end{minipage}
    
     \begin{minipage}[b]{.22\textwidth}
    \centering
        \includegraphics[width=\textwidth]{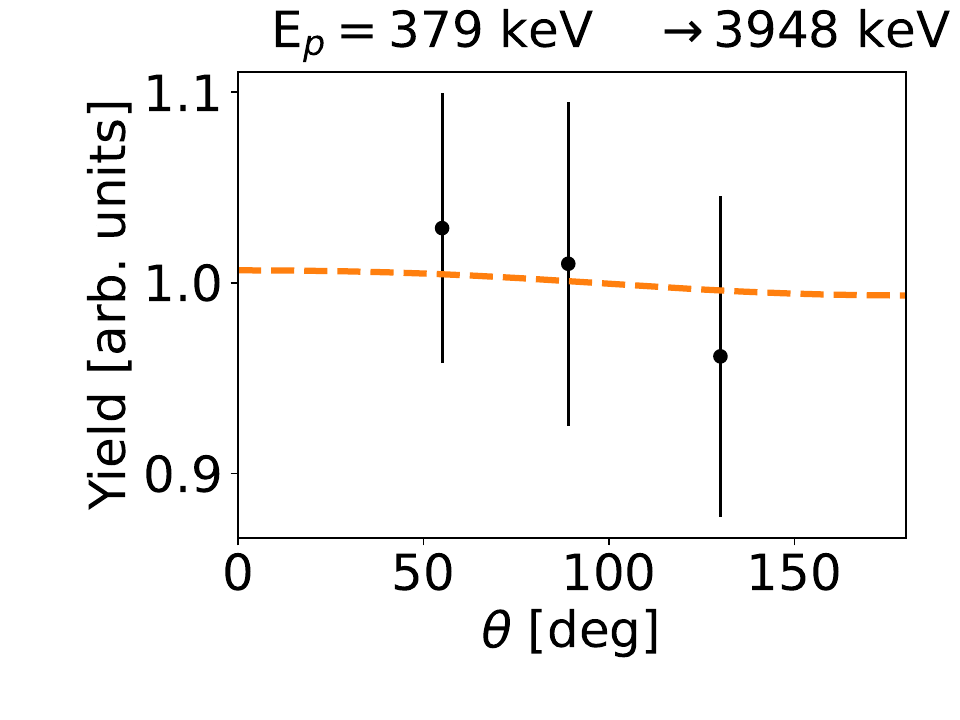}
     \end{minipage}
    \hfill
     \begin{minipage}[b]{.22\textwidth}
    \centering
        \includegraphics[width=\textwidth]{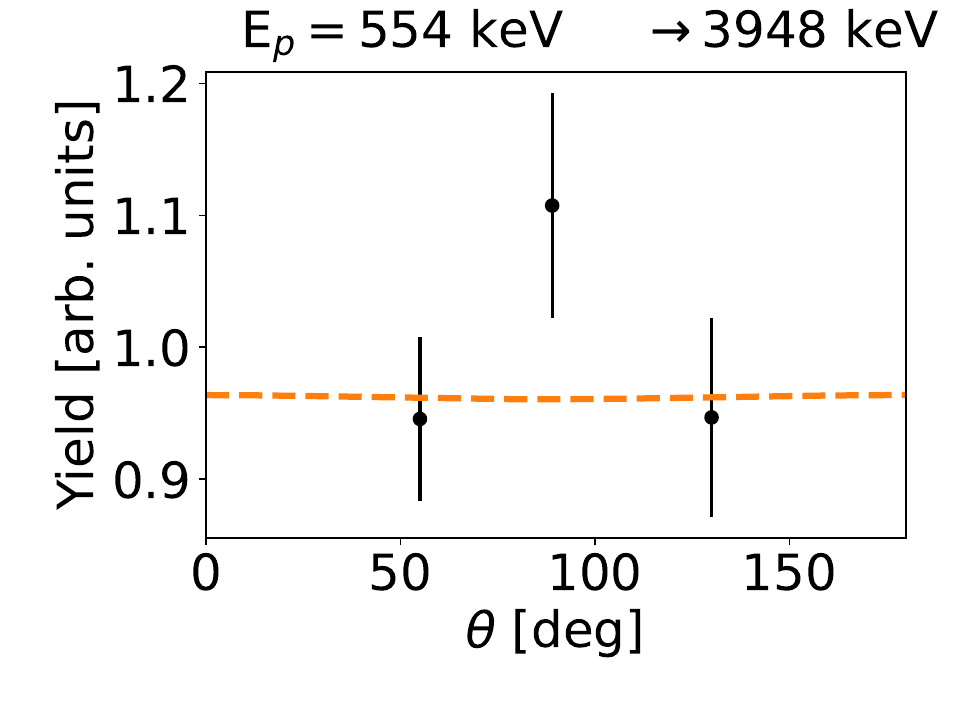}
     \end{minipage}
  \hfill
    \begin{minipage}[b]{.22\textwidth}
    \centering
        \includegraphics[width=\textwidth]{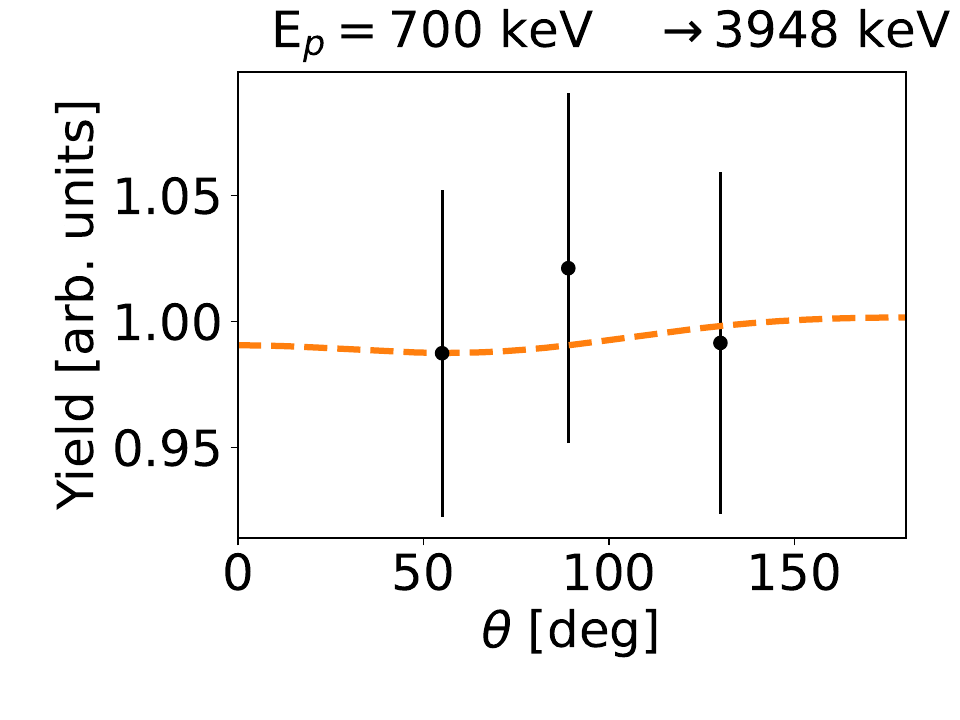}
     \end{minipage}
     \hfill
     \begin{minipage}[b]{.22\textwidth}
    \centering
        \includegraphics[width=\textwidth]{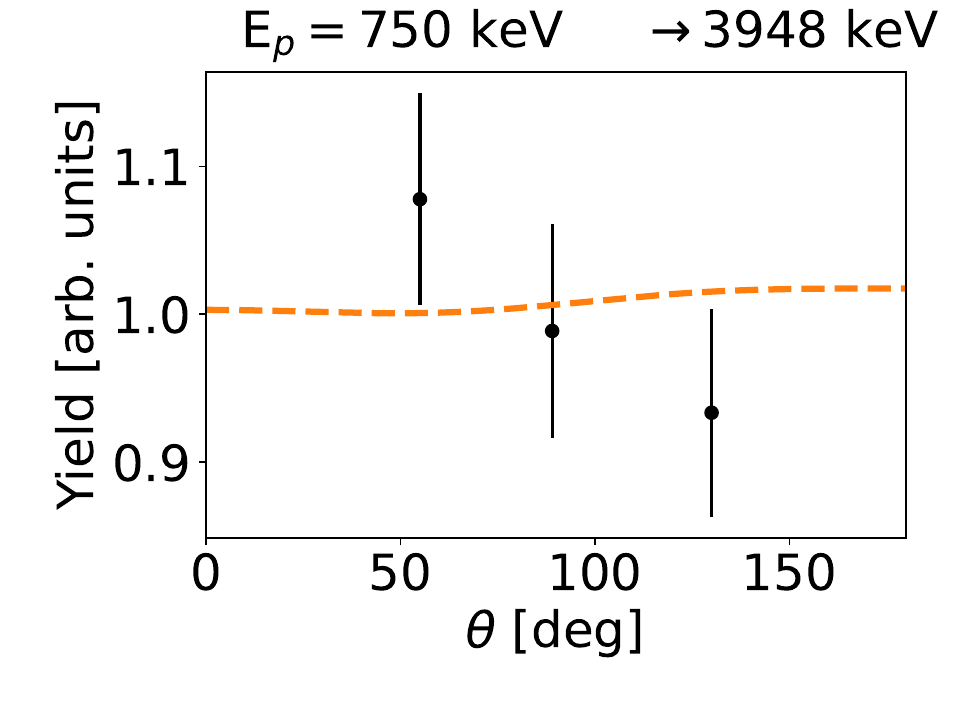}
     \end{minipage}
     
     \begin{minipage}[b]{.22\textwidth}
    \centering
        \includegraphics[width=\textwidth]{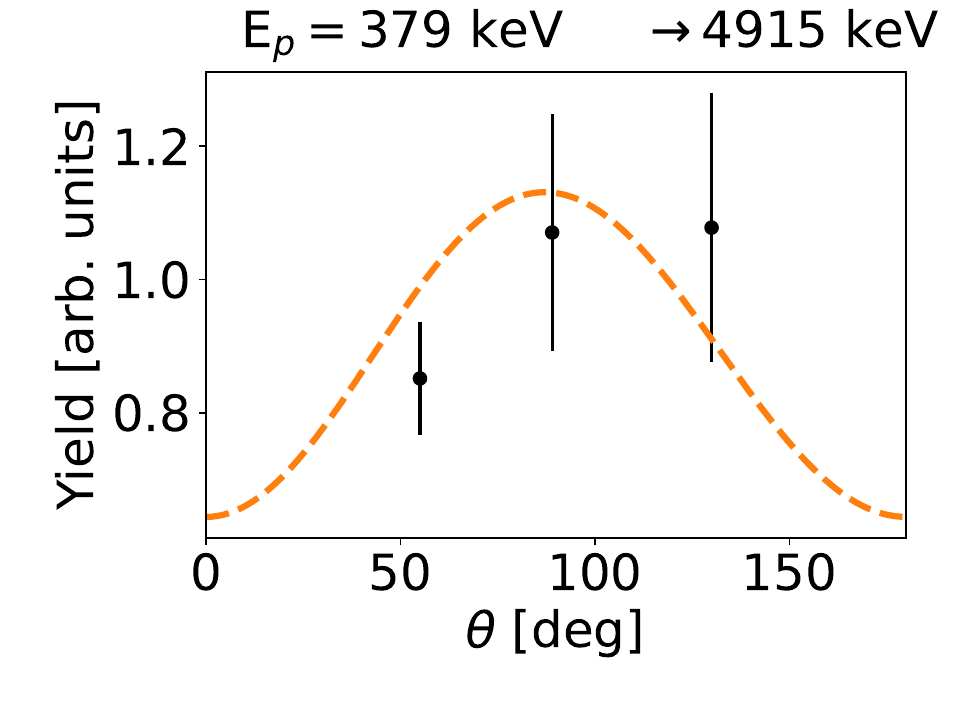}
     \end{minipage}
     \hfill
     \begin{minipage}[b]{.22\textwidth}
    \centering
        \includegraphics[width=\textwidth]{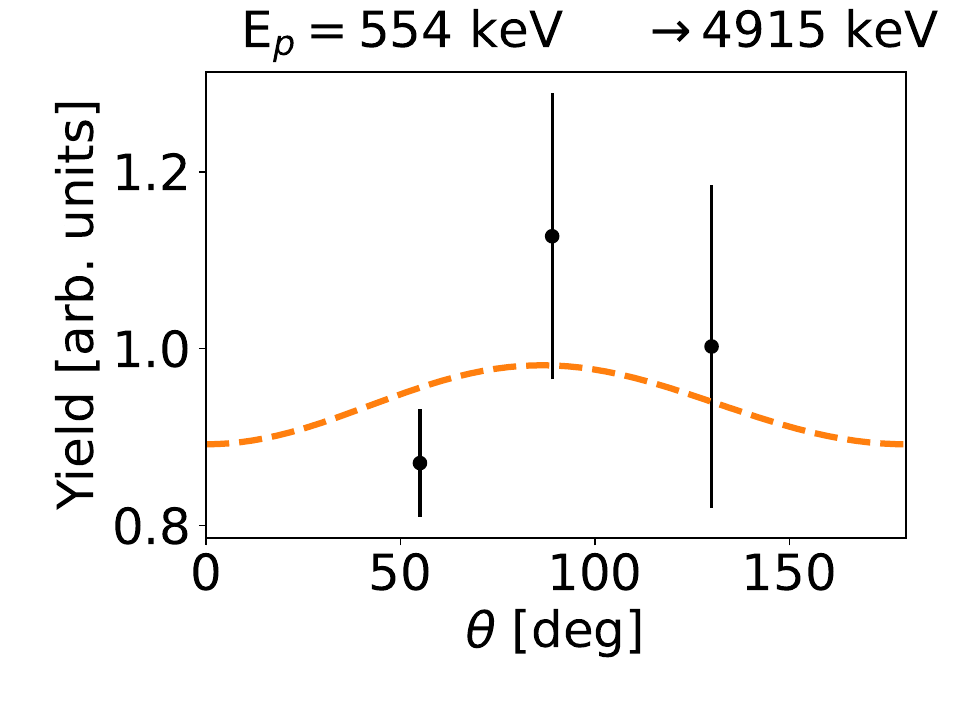}
     \end{minipage}
    \hfill
     \begin{minipage}[b]{.22\textwidth}
    \centering
        \includegraphics[width=\textwidth]{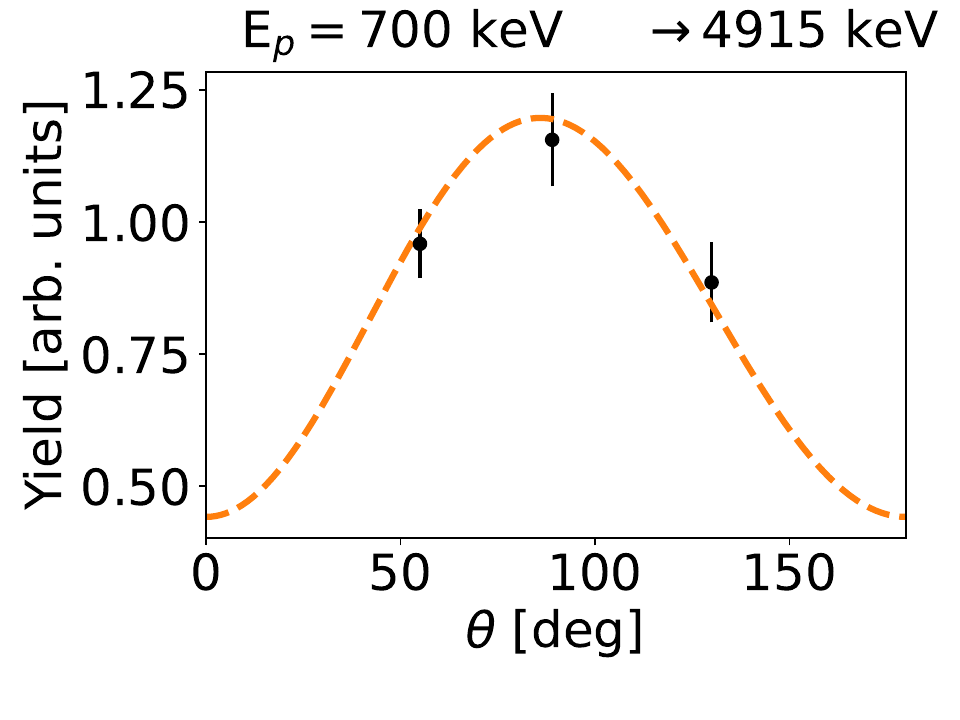}
     \end{minipage}
    \hfill
    \begin{minipage}[b]{.22\textwidth}
    \centering
        \includegraphics[width=\textwidth]{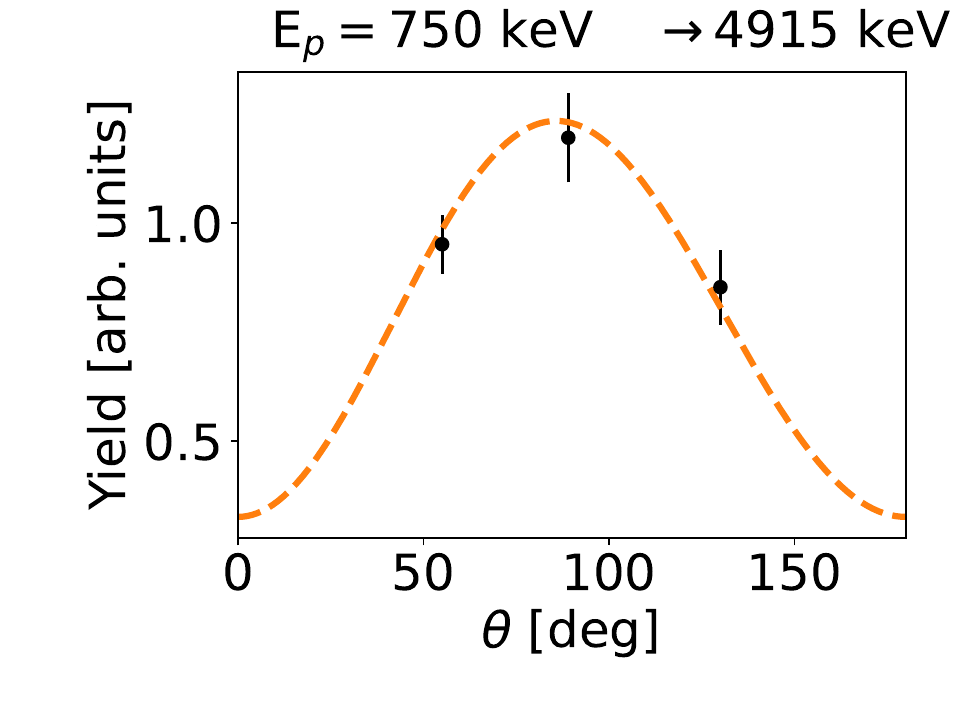}
     \end{minipage}

    \hfill
    \hfill
    \hfill
    \hfill
    \hfill
    \hfill
    \hfill
    \hfill
     \begin{minipage}[b]{.22\textwidth}
    \centering
        \includegraphics[width=\textwidth]{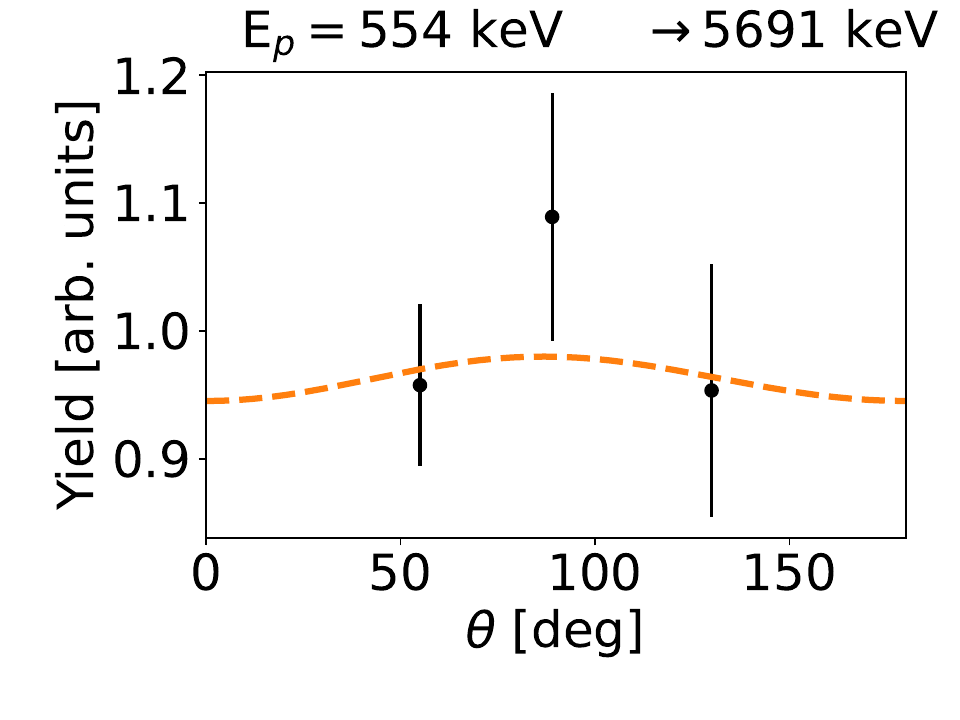}
     \end{minipage}
    \hfill
     \begin{minipage}[b]{.22\textwidth}
    \centering
        \includegraphics[width=\textwidth]{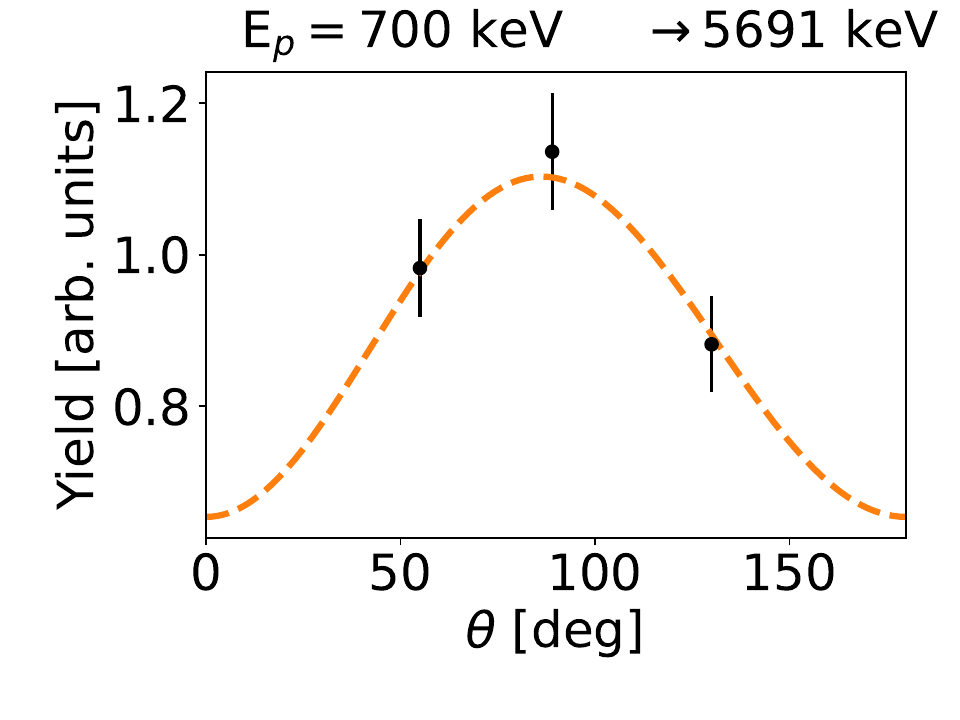}
     \end{minipage}
    \hfill
    \begin{minipage}[b]{.22\textwidth}
    \centering
        \includegraphics[width=\textwidth]{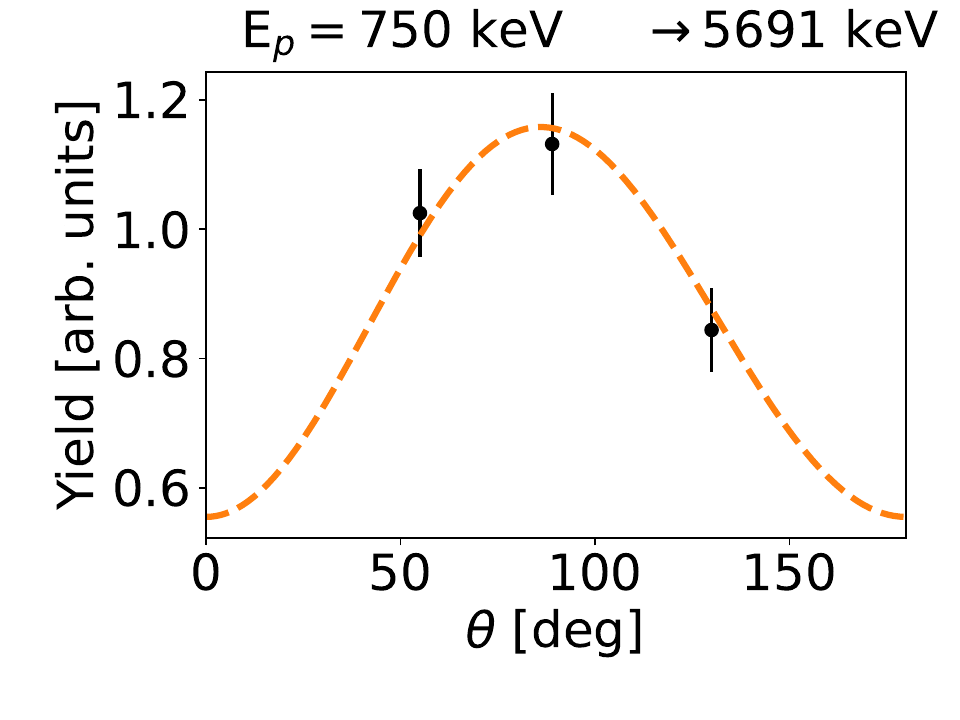}
    \end{minipage}
    
    \caption{The observed angular distributions, black dots, for different $^{13}$C($p,\gamma$)$^{14}$N transitions compared with extrapolation by Ref. \cite{chakraborty2015} of experimental data reported in Ref. \cite{king}, dashed line.
    The angle $\theta$ is in the laboratory frame. As reported in Ref. \cite{king} near $E$\textsubscript{p}$\,=\,$550~keV the anisotropies are compatible with unity. At energies far from the broad resonance, only the transition to \SI{4915}{\kilo\electronvolt} and \SI{5691}{\kilo\electronvolt} states show a non isotropic distribution, as found by Ref. \cite{king} and as predicted by the direct capture model in Ref. \cite{rolfs}. The plotted uncertainty was calculated as the sum in quadrature of  statistical (1$-$2\%) and the propagation of the efficiency uncertainty (\SI{6.5}{\percent}).}
    \label{fig:figure4}
\end{figure*}

At proton energies $E$\textsubscript{p} = 379, 554, 700 and 750~keV, the yields of all primary transitions, except to \SI{2312}{\kilo\electronvolt} and \SI{5105}{\kilo\electronvolt} states (due to low statistics), were calculated for detectors B, C and A in far geometry. The angular distribution results are shown in Fig. \ref{fig:figure4}. To correct for the finite volume of the detectors, the geometric $Q$ coefficients were used \cite{rose1953}. The calculated values lied between 0.98 and 1. Since present data are in agreement with the extrapolation by Ref. \cite{chakraborty2015} of literature angular distribution results \cite{king}, the latter was used to calculate the angular distribution correction, $W_i(\theta)$, in eq. \ref{eq_1}.

The $S$-factor, $S(E)$, was calculated using \cite{Rolfs88-Book}:

\begin{equation}
Y = \int {E^{-1} S(E) e^{-2\pi\eta(E)} \epsilon_{\mathrm{eff}}^{-1}(E)P(E) dE}
\label{eq_1}
\end{equation}

where $E$ is the proton beam energy in the center of mass frame, and $\eta(E)$ is the Sommerfeld parameter \cite{Rolfs88-Book}. The effective stopping power, $\epsilon_{\mathrm{eff}}(E)$, was calculated for the isotopic composition of present targets (99\% $^{13}$C and 1\% $^{12}$C) using SRIM database \cite{srim2003}. The effective stopping power uncertainty is assumed to be \SI{3.5}{\percent} as follows from experimental values reported in SRIM in the energy range of interest here. The target profile, $P(E)$, was obtained from peak-shape analysis \cite{Ciani20-EPJA}. In order to account for the $S$-factor energy dependence, the curve from the R-matrix evaluation in Ref. \cite{Skowronski-2023} was used. The effective energy, $E_{\textup{eff}}$, to be associated to each data point was estimated as defined in Ref. \cite{Brune-2013}.

For the narrow resonance at \SI{448.5}{\kilo\electronvolt}, the only scan that was performed during the measurement was analyzed by fitting the data points against eq. \ref{eq_1}. To calculate the contribution to the yield by the direct capture and the broad resonance component, the $S(E)$ curve from Ref. \cite{Skowronski-2023} was adopted. The target profile was assumed to have two steps, as in the peak shape case, and was left free to vary during the fit. The only fixed parameter in the target profile was the resonance energy, which was taken from Ref. \cite{Ajzenberg-Selove_1991NuPhA, vogl}. In order to perform the fit and to take into account the energy straggling of the incoming particle, the approach reported in Ref. \cite{manteigas2021} was used, leaving the $\omega \gamma$ of the resonance free to vary. Additionally, the branching ratios from Ref. \cite{king} were assumed due to insufficient statistics for this particular resonance. 

\section{\label{sec:results}Results}
The result for the 448.5 keV resonance is shown in Fig.\ref{fig:figure2}. The best fit indicates a new value of the total resonance strength, $\omega \gamma$ = 18(2) meV, to be compared with 21(2)meV \cite{king}. The present uncertainty is the total, 6\% statistical and 12\% systematic, while in Ref. \cite{king} there are no details on the nature of the uncertainty.

\begin{figure*}[htbp]
    \centering
    \begin{minipage}[b]{\textwidth}
        \includegraphics[width=0.8\textwidth]{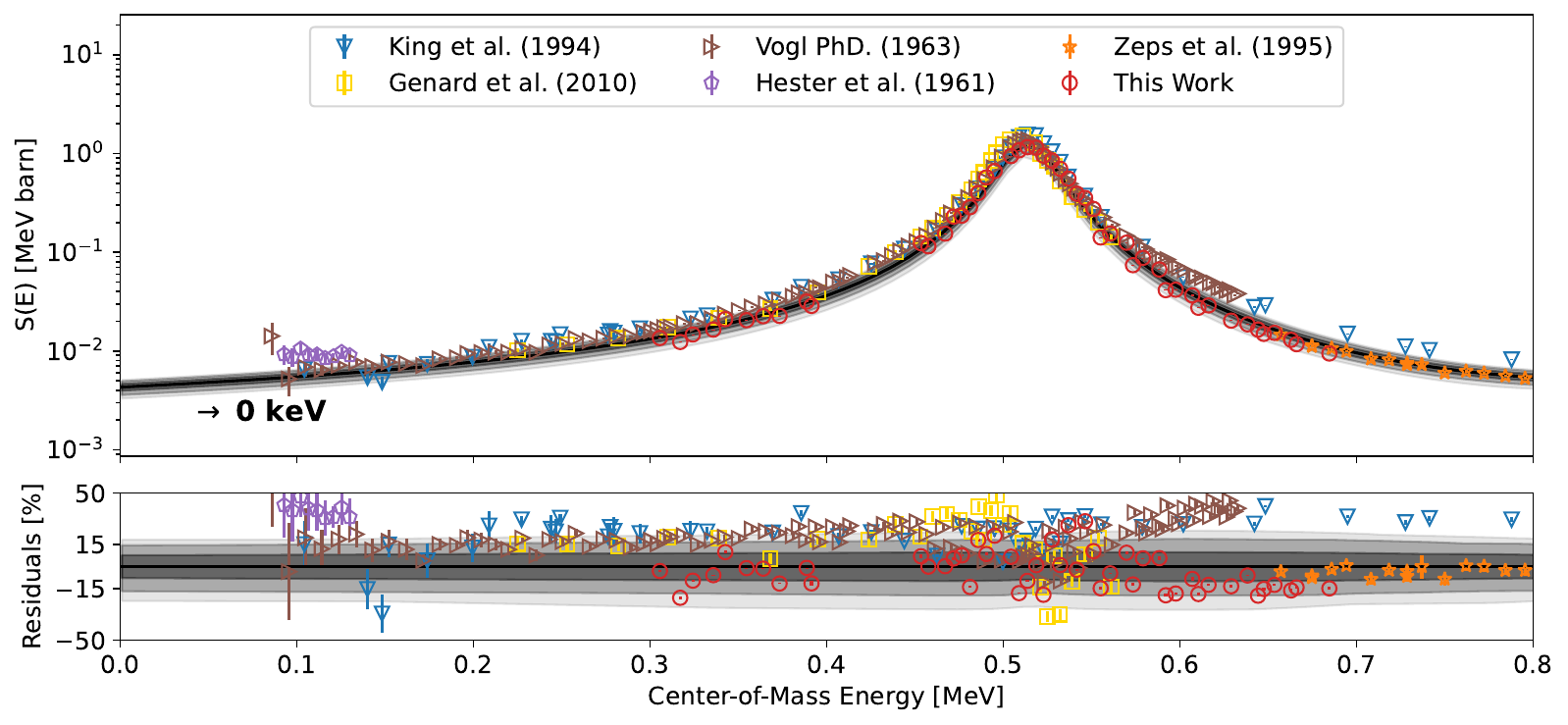}
    \end{minipage}
    
    \begin{minipage}[b]{0.49\textwidth}
        \includegraphics[width=0.9\textwidth]{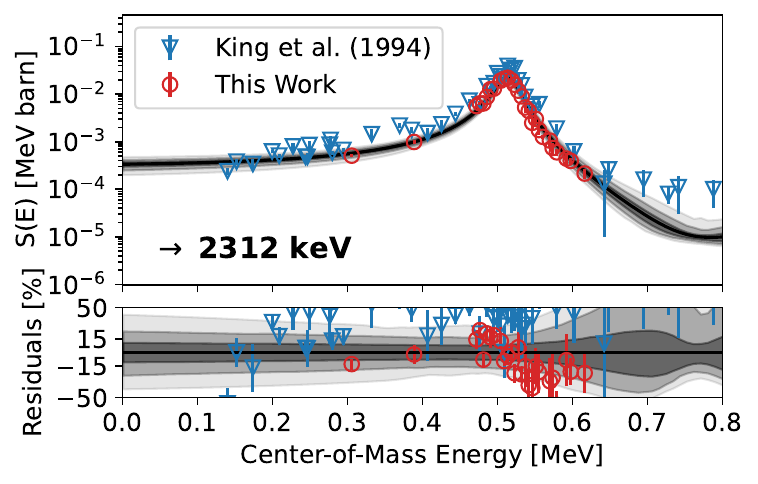}
    \end{minipage}
    \hfill
    \begin{minipage}[b]{0.49\textwidth}
        \includegraphics[width=0.9\textwidth]{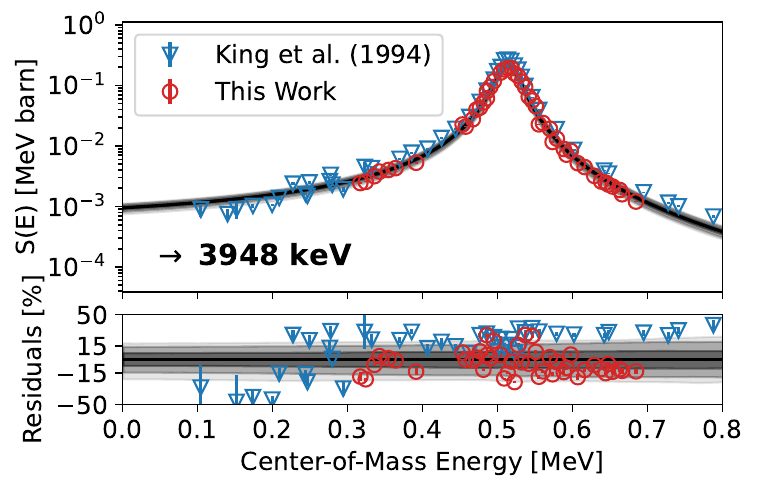}
    \end{minipage}

    \begin{minipage}[b]{0.49\textwidth}
        \includegraphics[width=0.9\textwidth]{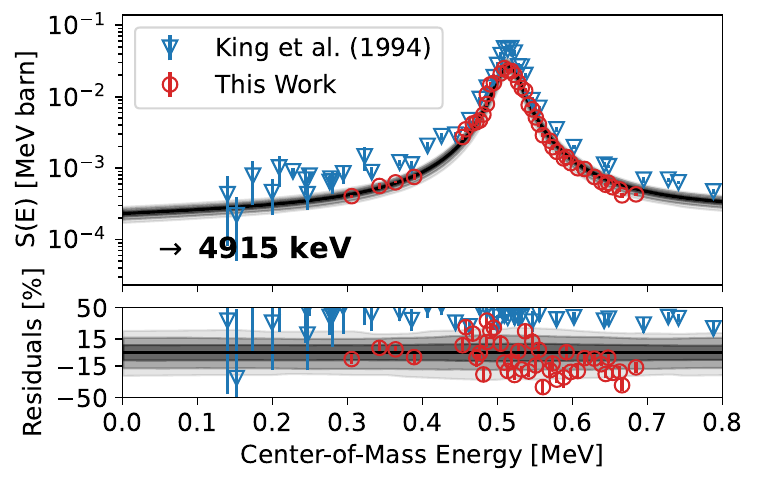}
    \end{minipage}
    \hfill
    \begin{minipage}[b]{0.49\textwidth}
        \includegraphics[width=0.9\textwidth]{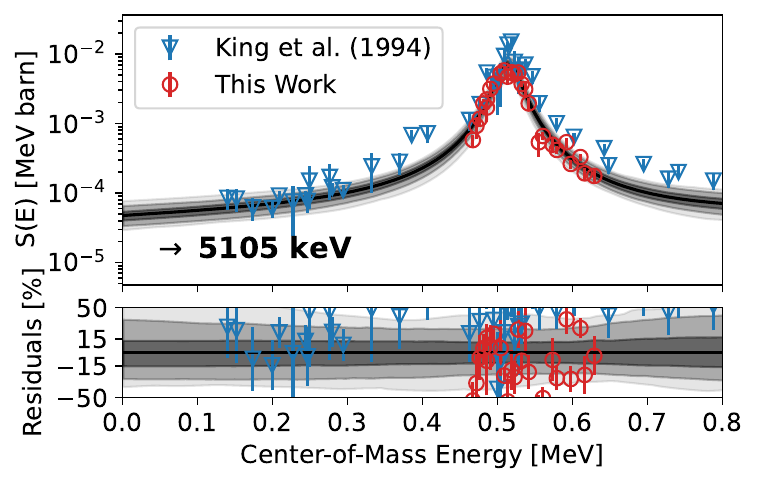}
    \end{minipage}

    \begin{minipage}[b]{0.49\textwidth}
        \includegraphics[width=0.9\textwidth]{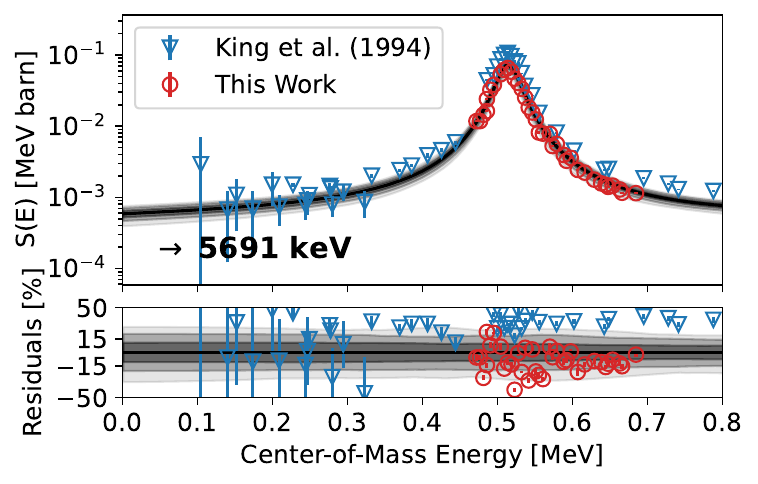}
    \end{minipage}
    \hfill
    \begin{minipage}[b]{0.49\textwidth}
        \includegraphics[width=0.9\textwidth]{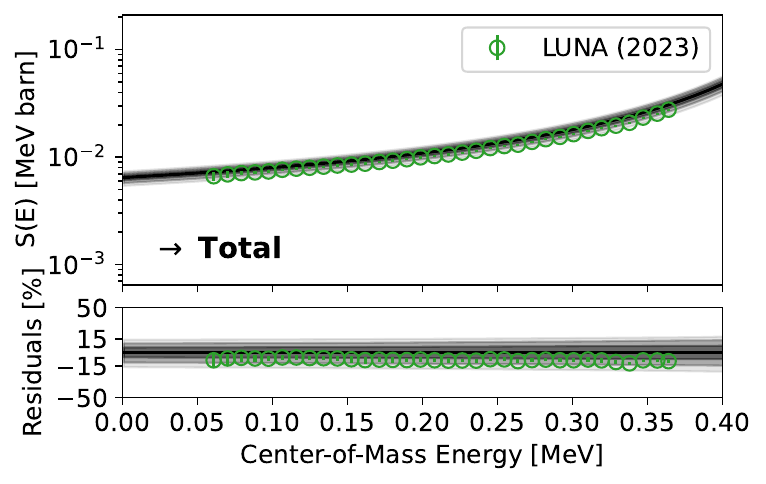}
    \end{minipage}
    
    \caption{The result of the R-matrix fit, black line, of the $^{13}$C($p,\gamma$)$^{14}$N reaction cross section data available in literature \cite{Skowronski-2023, king, genard, vogl, hester, zeps1995}. The black shaded areas show $1\sigma$, $2\sigma$ and $3\sigma$ uncertainties. The data in green for the transitions to the excited states and the total $S$-factor are from Ref. \cite{Skowronski-PhD}.}
    \label{fig:figure5}
\end{figure*}

The new partial results for the $^{13}$C(p,$\gamma$)$^{14}$N reaction $S(E)$ are shown in Fig.\ref{fig:figure5} with only statistical uncertainties.

The systematic uncertainty amounts to 12\% including the contributions detailed in Table \ref{tab:table3}.

\begin{table}
  \centering
    \caption{Contributions to the overall systematic uncertainty in the $^{13}$C(p,$\gamma$)$^{14}$N $S$ factor arising from different sources.}
    \begin{tabular}{@{}ll@{}}
        \midrule
        \midrule
        Source & $\Delta S/S$ [\%] \\
        \midrule
         Efficiency & 11 \\
         stopping power & 3.5 \\
          target profile & 3 \\
         charge collection & 3 \\
         angular distribution & $<\,2$ \\
        \midrule
        Total & 12\\
        \midrule
        \midrule
    \end{tabular}
    \label{tab:table3}
\end{table}

Over the whole energy range, a 20\% scaling difference is observed between present results and literature data for the main transition \cite{vogl, king, genard}. At low energy the present data are consistent with recent results reported by the LUNA collaboration \cite{Skowronski-2023}.

An R-matrix analysis was performed using AZURE2 code \cite{azuma2010} and considering all the data available. Both the elastic \cite{hebbard1960} and radiative capture channels were considered. For the latter the data from \cite{Skowronski-2023,king,vogl,genard,hester,zeps1995} and present work were used.
The channel radius was set to $4.3$ fm. All known excited states of the $^{14}$N nucleus up to \SI{10.5}{\mega\electronvolt} \cite{Ajzenberg-Selove_1991NuPhA} were included in the calculation. Additionally, two background poles with $0^{-}$ and $1^{-}$ spin-parities, respectively, were included and arbitrarily put at \SI{20}{\mega\electronvolt}. The fit was done following the hybrid method reported in Ref. \cite{skowronski2025}, where for each iteration the normalization of every dataset was sampled assuming a log-normal distribution given by its systematic uncertainty, then fixed and fitted, with a total of 1000 samples. The only normalization factors left free during the fit are the ones for the datasets reported in Ref. \cite{zeps1995,vogl,hester}, since no systematic error is reported. Reference \cite{zeps1995} does not provide any absolute normalization of the data and Ref. \cite{vogl,hester} do not report any systematic uncertainty. During the fit, only ANCs to the subthreshold states and the parameters of the \SI{8.062}{\mega\electronvolt} and \SI{8.802}{\mega\electronvolt} states, and the respective background poles, were left free to vary in the fit. For all the other resonances, the parameters from Ref. \cite{zeps1995} were used. Additionally, the ANCs were treated as nuisance parameters with the reference values taken from Ref. \cite{mukhamedzhanov2003}. The R-matrix curve is in Fig.~\ref{fig:figure5}, and the best-fit resonance parameters are reported in Tab.~\ref{tab:table1}.

The present resonance energy is in agreement with experimental results in Ref. \cite{vogl, king} and recent evaluation of the the $^{13}$C(p,$\gamma$)$^{14}$N reaction cross section \cite{chakraborty2015, Li-2012, Artemov-2008, Mukhamedzhanov-2003}.
On the other hand, the present proton width is lower of about 6\% with respect data in literature \cite{chakraborty2015, Li-2012, Artemov-2008}, however in agreement with the result in Ref. \cite{Li-2012}, given its high uncertainty.
The radiative width is 22\% lower than results reported in literature \cite{chakraborty2015, Li-2012, Artemov-2008, Mukhamedzhanov-2003, Ajzenberg-Selove_1991NuPhA}, mainly because of the present and LUNA results. 

The present R-matrix analysis provides a new extrapolation of the cross section down to zero energy, $S$\textsubscript{tot}(0) = 6.4(4) keV b which is in agreement with the recent evaluation by Ref. \cite{chakraborty2015} and it is consistent with previous reported results \cite{xu2013, Mukhamedzhanov-2003, nacre, king, genard, vogl, fowler-c12, hebbard1960}, but has higher precision, see Tab.\ref{tab:table4}. Only exception is the result reported in Ref. \cite{Li-2012}, which is higher than the present by more than 3$\sigma$.

\setcitestyle{authoryear}
\begin{table}[htbp]
    \centering
    \caption{The $S$\textsubscript{tot}(0) results from the present R-matrix fit and from literature.}
    \begin{tabular}{@{}ll@{}}
        \midrule
        \midrule
        Reference & $S$\textsubscript{tot}(0) [keV b] \\
        \midrule
         This work & 6.4(4) \\
         \citet{chakraborty2015} & 6.83(95) \\
          \citet{xu2013} & 8.1(12) \\
         \citet{Li-2012} & 7.92(49) \\
         \citet{genard} & 4.85(76) \\
        \citet{Artemov-2008} & 6.944 \\
        \citet{Mukhamedzhanov-2003} & 7.58(110) \\
        \citet{nacre} & 7.0(15) \\
        \citet{king} & 7.6(10) \\
        \citet{hebbard1960} & 6.0(8)\\
       \citet{vogl} & 5.5(8)\\
      \citet{fowler-c12} & 5.9\\
        \midrule
        \midrule
    \end{tabular}
    \label{tab:table4}
\end{table}
\setcitestyle{authoryear}

\setcitestyle{numbers}

\begin{figure}[!h]
\includegraphics[width=0.5\textwidth]{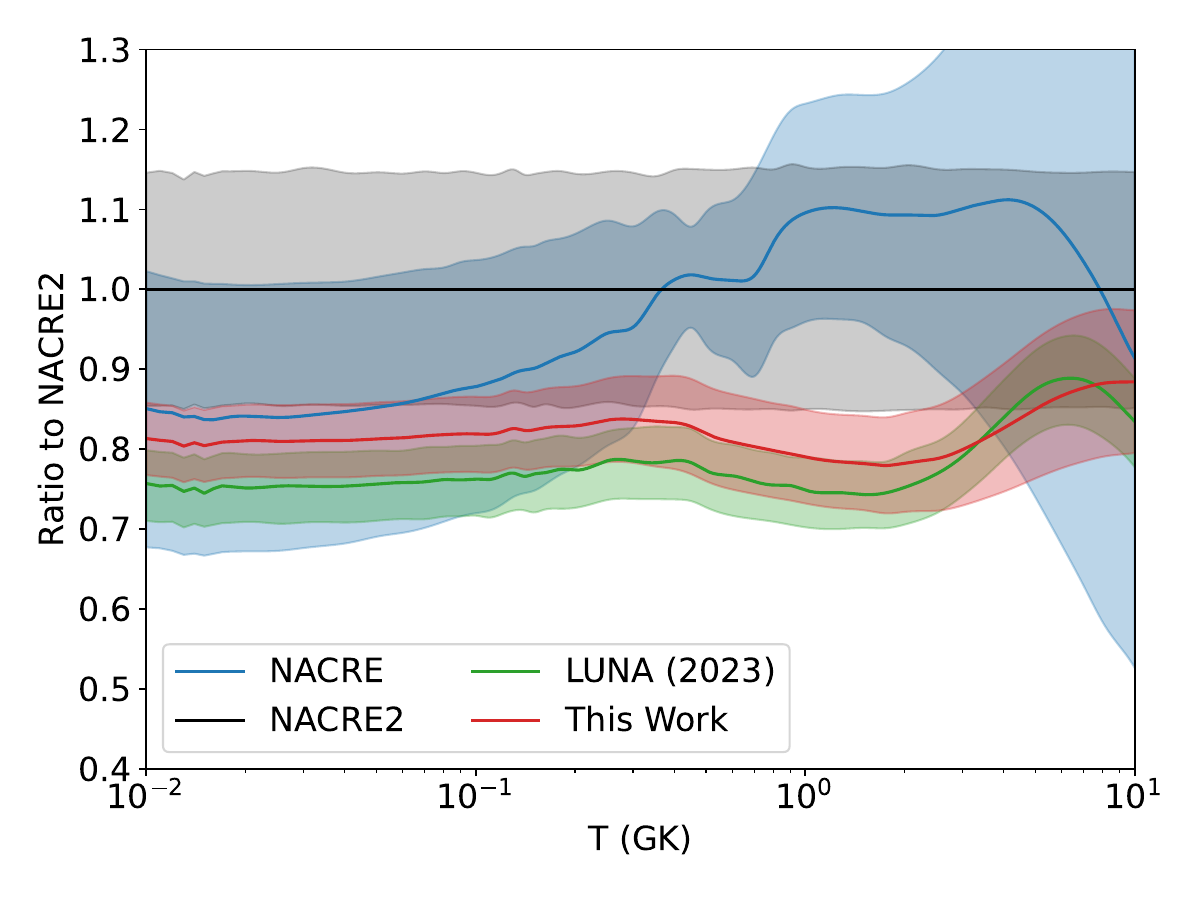}
\caption{The astrophysical reaction rate from the present work (red), normalized to NACRE2 \cite{xu2013}, black line. The uncertainty of the present rate is $7$\% over the entire temperature range. The reaction rate by Ref. \cite{Skowronski-2023,nacre} are reported for comparison in green and blue, respectively.}
\label{fig:figure6}
\end{figure}

To evaluate the impact of the present results, the thermonuclear reaction rate was calculated with the present R-matrix fit of the $S$-factor,
and compared with the most widely adopted reaction rates \cite{nacre,xu2013} and the recent revision by LUNA \cite{Skowronski-2023}, see Fig.\ref{fig:figure6}.

The present reaction rate uncertainty is significantly reduced compared to both Ref. \cite{nacre, xu2013} over the whole temperature range, 0.01-10~GK. Up to 10 GK the present rate is consistent with the one reported in Ref. \cite{Skowronski-2023}, while it significantly differs from Ref. \cite{xu2013}.

\section{\label{sec:conclusion}Conclusion}
The $^{13}$C(p,$\gamma$)$^{14}$N partial and total cross-section has been measured in a wide energy range, 350$<E$\textsubscript{p}$<$750~keV, at the Felsenkeller facility, with a total uncertainty of about 12\%.
The present $S(E)$ results are about 20\% lower than data available in the literature over the whole energy range explored. But they are, consistent with recent results by LUNA at low energies \cite{Skowronski-2023}.
A comprehensive R-matrix analysis has been performed to derive new precise parameters for the broad resonance at $E$\textsubscript{p} = 551 keV. On the other hand for the $E$\textsubscript{p} = 448.5 keV narrow resonance a new value for the resonance strength is reported.
Finally, the calculated reaction rate is consistently lower than literature, up to 10 GK suggesting a revision of the stellar model calculations and the need for a renewed evaluation of the impact on CNO nucleosynthesis.

\paragraph{Acknowledgments $-$}
The authors would like Gy. Gy\"urky (ATOMKI) for producing the targets.
Financial support by INFN, the
Italian Ministry of Education, University and Research (MIUR) through the ``Dipartimenti di eccellenza'' project ``Science of the Universe'', European Union, ChETEC-INFRA (grant agreement no. 101008324),  the Hungarian National Research, Development and
Innovation Office (NKFIH K134197 and FK134845),  are gratefully acknowledged. M. A. acknowledges funding by STFC UK (grant no. ST/L005824/1). D.R. acknowledges funding from the European Research Council (ERC) under grant agreement no. 852016. E.M. acknowledges an Alexander von Humboldt postdoctoral fellowship.
For the purpose of open access, the author has applied a Creative Commons Attribution (CC BY) licence to any Author Accepted Manuscript version arising from this submission.

\end{document}